\documentclass[lettersize]{emulateapj}

\usepackage{natbib}
\usepackage{amsmath}
\usepackage{multirow}
\usepackage{float}
\usepackage{color}
\usepackage{afterpage}
\usepackage[abs]{overpic}
\usepackage{CJK}
\usepackage{hyperref}
\usepackage{rotating}

\definecolor{Red}{rgb}{1,0,0}
\definecolor{Blue}{rgb}{0,0,1}
\definecolor{Green}{rgb}{0,1,0}
\definecolor{magenta}{rgb}{1,0,.6}
\definecolor{lightblue}{rgb}{0,.5,1}
\definecolor{lightpurple}{rgb}{.6,.4,1}
\definecolor{gold}{rgb}{.6,.5,0}
\definecolor{orange}{rgb}{1,0.4,0}
\definecolor{hotpink}{rgb}{1,0,0.5}
\definecolor{newcolor2}{rgb}{.5,.3,.5}
\definecolor{newcolor}{rgb}{0,.3,1}
\definecolor{newcolor3}{rgb}{1,0,.35}
\definecolor{darkgreen1}{rgb}{0, .35, 0}
\definecolor{darkgreen}{rgb}{0, .6, 0}
\definecolor{darkred}{rgb}{.75,0,0}

\newcommand{\jcap}{JCAP}

\def \arcmin     { ^{\prime} }
\def \arcsec    {^{\prime\prime}}

\def \mpc       {{\rm\ Mpc}}

\def \kms            {~{\rm km~s}^{-1}}

\def \etal      {et~al.\ }

\def \hMpc      {h^{-1}{\rm\ Mpc}}

\newcommand{\alphat}{\ensuremath{\alpha}}

\newcommand{\snr}{\ensuremath{\mathrm{S/N}}}
\newcommand{\lya}{Ly$\alpha$}

\newcommand{\waveion}[3]{\ion{#1}{#2} $\lambda$#3}

\newcommand{\beq}{\begin{equation}}
\newcommand{\eeq}{\end{equation}}
\newcommand{\bc}{\begin{center}}
\newcommand{\ec}{\end{center}}
\newcommand{\bfig}{\begin{figure}}
\newcommand{\efig}{\end{figure}}

\newcommand{\fmean}{\ensuremath{\langle F \rangle}}

\newcommand{\zalp}{\ensuremath{z_\alpha}}

\newcommand{\zbg}{\ensuremath{z_\mathrm{bg}}}

\newcommand{\ang}{\ensuremath{\mathrm{\AA}}}

\newcommand{\dperp}{\ensuremath{\langle d_{\perp} \rangle}}
\newcommand{\sigthreed}{\ensuremath{\epsilon_{\rm 3D}}}

\newcommand{\delrecon}{\ensuremath{\delta_F^{\rm rec}}}

\newcommand{\persqdeg}{\mathrm{deg}^{-2}}
\newcommand{\sqdeg}{\mathrm{deg}^{2}}
\newcommand{\cmd}{\mathbf{C}_\mathrm{MD}}
\newcommand{\cdd}{\mathbf{C}_\mathrm{DD}}

\newcommand{\yperp}{\ensuremath{y_\mathrm{perp}}}
\newcommand{\xperp}{\ensuremath{x_\mathrm{perp}}}

\def \etal {et~al.}

\slugcomment{Accepted for ApJL}
\shorttitle{\lya\ Forest Tomography with $z\sim 2-3$ Galaxies}
\shortauthors{Lee \etal}

\begin{document}

\title{Ly$\alpha$ Forest Tomography from Background Galaxies: \\
The First Megaparsec-Resolution Large-Scale Structure Map at $\lowercase{z}> 2$}
\author{Khee-Gan Lee\altaffilmark{1}, 
Joseph F. Hennawi\altaffilmark{1}, 
Casey Stark\altaffilmark{2,3},
J.\ Xavier Prochaska\altaffilmark{4,5}, 
Martin White\altaffilmark{2,3}, \\
David J. Schlegel\altaffilmark{5},
Anna-Christina Eilers\altaffilmark{1},
Andreu Arinyo-i-Prats\altaffilmark{6}, 
Nao Suzuki\altaffilmark{7},
Rupert~A.C.~Croft\altaffilmark{8},  \\
Karina I. Caputi\altaffilmark{9}, 
Paolo Cassata\altaffilmark{10,11},
Olivier Ilbert\altaffilmark{11},
Bianca Garilli\altaffilmark{12}, 
Anton M. Koekemoer\altaffilmark{13},\\
Vincent Le Brun\altaffilmark{11},
Olivier Le F\`evre\altaffilmark{11}, 
Dario Maccagni\altaffilmark{15}, 
Peter Nugent\altaffilmark{3,2},
Yoshiaki Taniguchi\altaffilmark{14},\\
Lidia A.M. Tasca\altaffilmark{11},
Laurence Tresse\altaffilmark{11},
Gianni Zamorani\altaffilmark{15},
Elena Zucca\altaffilmark{15}
}
\altaffiltext{1}{Max Planck Institute for Astronomy, K\"{o}nigstuhl 17, D-69117 Heidelberg, West Germany}
\altaffiltext{2}{Department of Astronomy, University of California at Berkeley, B-20 Hearst Field Annex \# 3411,
Berkeley, CA 94720, USA}
\altaffiltext{3}{Lawrence Berkeley National Laboratory, 1 Cyclotron Rd., Berkeley, CA 94720, USA}
\altaffiltext{4}{Department of Astronomy and Astrophysics, University of California, 1156 High Street, Santa Cruz, CA 95064, USA}
\altaffiltext{5}{University of California Observatories, Lick Observatory 1156 High Street, Santa Cruz, CA 95064, USA}
\altaffiltext{6}{Institut de Ci\`encies del Cosmos, Universitat de Barcelona (IEEC-UB), Mart\'i Franqu\`es 1, E08028 Barcelona, Spain}
\altaffiltext{7}{Kavli Institute for the Physics and Mathematics of the Universe (IPMU), The University of Tokyo, 
Kashiwano-ha 5-1-5, Kashiwa-shi, Chiba, Japan}
\altaffiltext{8}{Department of Physics, Carnegie-Mellon University, 5000 Forbes Avenue, Pittsburgh, PA 15213, USA}
\altaffiltext{9}{Kapteyn Astronomical Institute, University of Groningen, P.O. Box 800, 9700 AV Groningen, The Netherlands}
\altaffiltext{10}{Instituto de Fisica y Astronomia, Facultad de Ciencias, Universidad de Valparaiso, Av. Gran Bretana 1111, Casilla 5030, Valparaiso, Chile}
\altaffiltext{11}{Aix Marseille Universit\'e, CNRS, LAM (Laboratoire d'Astrophysique  de Marseille) UMR 7326, 13388, Marseille, France}
\altaffiltext{12}{INAF--IASF, via Bassini 15, I-20133,  Milano, Italy}
\altaffiltext{13}{Space Telescope Science Institute, 3700 San Martin Drive, Baltimore MD 21218, USA}
\altaffiltext{14}{Research Center for Space and Cosmic Evolution, Ehime University, 2-5 Bunkyo-cho, Matsuyama 790-8577, Japan}
\altaffiltext{15}{INAF--Osservatorio Astronomico di Bologna, via Ranzani,1, I-40127, Bologna, Italy}
\email{lee@mpia.de}

%\begin{abstract}
%We present the first observations of foreground Lyman-$\alpha$ forest
%absorption from high-redshift galaxies, targeting 24 star-forming
%galaxies (SFGs) with $z\sim 2.3-2.8$ within a
%$5\arcmin\times15\arcmin$ region of the COSMOS field.  The transverse
%sightline separation is $\sim 2\,\hMpc$ comoving, allowing us to
%create a tomographic reconstruction of the 3D \lya\ forest absorption
%field over the redshift range $2.20\leq z\leq 2.45$.  The resulting
%map covers $6\,\hMpc\times 14\,\hMpc$ in the transverse plane and
%$230\,\hMpc$ along the line-of-sight with a spatial resolution of
%$\approx 3.5\,\hMpc$, and is the first time large-scale structure has
%been mapped on $\sim\mathrm{Mpc}$ scales at $z>2$.  We see significant
%structures extending across $\gtrsim 10\,\hMpc$ in the map, including
%several spanning the entire transverse breadth.  Simulated
%reconstructions with the same sightline sampling and spectral
%signal-to-noise show a good recovery of the underlying 3D absorption
%field.  Using data from other surveys, we identified 18 galaxies with
%known redshifts coeval with our map volume, enabling a direct
%comparison which shows that the galaxies preferentially occupy
%high-density regions within the tomographic map. This result
%establishes the feasbility of the CLAMATO survey, which aims to obtain
%\lya\ forest spectra for $\sim 1000$ SFGs over $\sim1 \,\sqdeg$ of the
%COSMOS field, in order to map out IGM large-scale structure at
%$\langle z \rangle \sim 2.3$ over a large volume $(100\,\hMpc)^3$.
%\end{abstract}

\begin{abstract}
We present the first observations of foreground Lyman-$\alpha$ forest absorption from high-redshift galaxies, 
targeting 24 star-forming galaxies (SFGs)  with $z\sim 2.3-2.8$
within a $5\arcmin\times15\arcmin$ region of the COSMOS field.
The transverse sightline separation is $\sim 2\,\hMpc$ 
comoving, allowing us to create a tomographic reconstruction of the 3D \lya\ forest absorption field over the redshift range $2.20\leq z\leq 2.45$. 
The resulting map covers $6\,\hMpc\times 14\,\hMpc$ in the transverse plane and $230\,\hMpc$ along 
the line-of-sight with a spatial resolution of $\approx 3.5\,\hMpc$, and is the first high-fidelity map of large-scale structure on 
$\sim\mathrm{Mpc}$ scales at $z>2$.
Our map reveals significant structures with $\gtrsim 10\,\hMpc$
extent, including several spanning the entire transverse breadth,
providing qualitative evidence for the filamentary structures
predicted to exist in the high-redshift cosmic web. Simulated reconstructions with
the same sightline sampling, spectral resolution, and signal-to-noise
ratio recover the salient structures present in the underlying 3D
absorption fields.  Using data from other surveys, we identified 18
galaxies with known redshifts coeval with our map volume enabling a
direct comparison to our tomographic map. This shows that galaxies
preferentially occupy high-density regions, in qualitative agreement
with the same comparison applied to simulations. Our results
establishes the feasibility of the CLAMATO survey, which aims to obtain
\lya\ forest spectra for $\sim 1000$ SFGs over $\sim1 \,\sqdeg$ of the
COSMOS field, in order to map out IGM large-scale structure at
$\langle z \rangle \sim 2.3$ over a large volume $(100\,\hMpc)^3$.
\end{abstract}


\section{Introduction}
The Lyman-\alphat\ (\lya) `forest' absorption seen in quasar spectra is a crucial probe of the
intergalactic medium (IGM). In the modern
`fluctuating Gunn-Peterson' scenario \citep{cen:1994,bi:1995,croft:1998,hui:1997}, this is from residual neutral hydrogen in 
photoionization-equilibrium, tracing the underlying density field, allowing the study of
 large-scale structure (LSS) at $z\gtrsim 2$ \citep[e.g.,][]{croft:2002,mcdonald:2006,busca:2013,palanque-delabrouille:2013a,
 delubac:2014}.

The \lya\ forest observed in individual quasars probe the IGM along 1-dimension, but 
using multiple spectra with small transverse separations, 
it is possible to `tomographically' reconstruct a 3D map of the \lya\
forest absorption (\citealt{pichon:2001}; \citealt{caucci:2008}; \citealt{cisewski:2014}; \citealt{lee:2014}, hereafter L14).  
The effective spatial-resolution, \sigthreed, of such a map is determined by the transverse sightline separation, \dperp.
This probes $\sim\mpc$ scales only by exploiting UV-bright star-forming galaxies (SFGs)
 as background sources in addition to quasars. However, SFGs are faint ($g\gtrsim 23$) --- 
even with 8-10m telescopes, only spectral S/N
of a few are feasible from such objects, assuming reasonable exposure times.
However, L14 argued that such data at moderate resolutions ($R\equiv\lambda/\Delta(\lambda)\sim 1000$) 
are adequate for \lya\ forest tomography that resolve
the LSS on scales of $\sigthreed\sim 2-5\,\hMpc$.

In this Letter, we describe pilot observations for the COSMOS Lyman-Alpha Mapping And Tomography Observations 
(CLAMATO) survey. The full survey is aimed at mapping the $z\sim2.3$ IGM
within $\sim1\,\sqdeg$ of the 
COSMOS field \citep{scoville:2007}. 
The pilot observations were however limited to one half-night of successful data, yielding moderate-resolution
 spectra for 24 SFGs at $g\leq24.9$ within $\sim5\arcmin\times14\arcmin$.
 
 This data represents, to our knowledge, the first systematic attempt to exploit spectra of unlensed high-redshift SFGs
  for \lya\ forest analysis. 
 Our background sources are $\sim2.5-3$mag fainter than existing \lya\ forest datasets 
 \citep[e.g., $g\sim21.5$ in BOSS,][]{dawson:2013}, yielding $\sim 100$ greater area density of sightlines 
($\sim1000\,\persqdeg$ vs $\sim15\,\persqdeg$ in BOSS). 
This dramatic increase results in small average inter-sightline separations ($\dperp\sim 2.3\,\hMpc$), enabling a  
tomographic reconstruction of the 3D \lya\ forest absorption, providing
an unprecedented view of the $z>2$ cosmic web on scales of several comoving Mpc. 
As we shall see, comparisons with a small number of coeval galaxies 
as well as simulated reconstructions
indicate that the map is indeed tracing LSS.
 
 In this paper, we assume a concordance flat $\Lambda$CDM cosmology with $\Omega_M=0.26$, $\Omega_\Lambda=0.74$,
 and $H_0=70\,\kms$.
 
\section{Observations and Data Reduction}
\begin{figure*}[ht]\begin{center}\includegraphics[height=0.73\textheight,clip=true,trim=23 20 33 30]{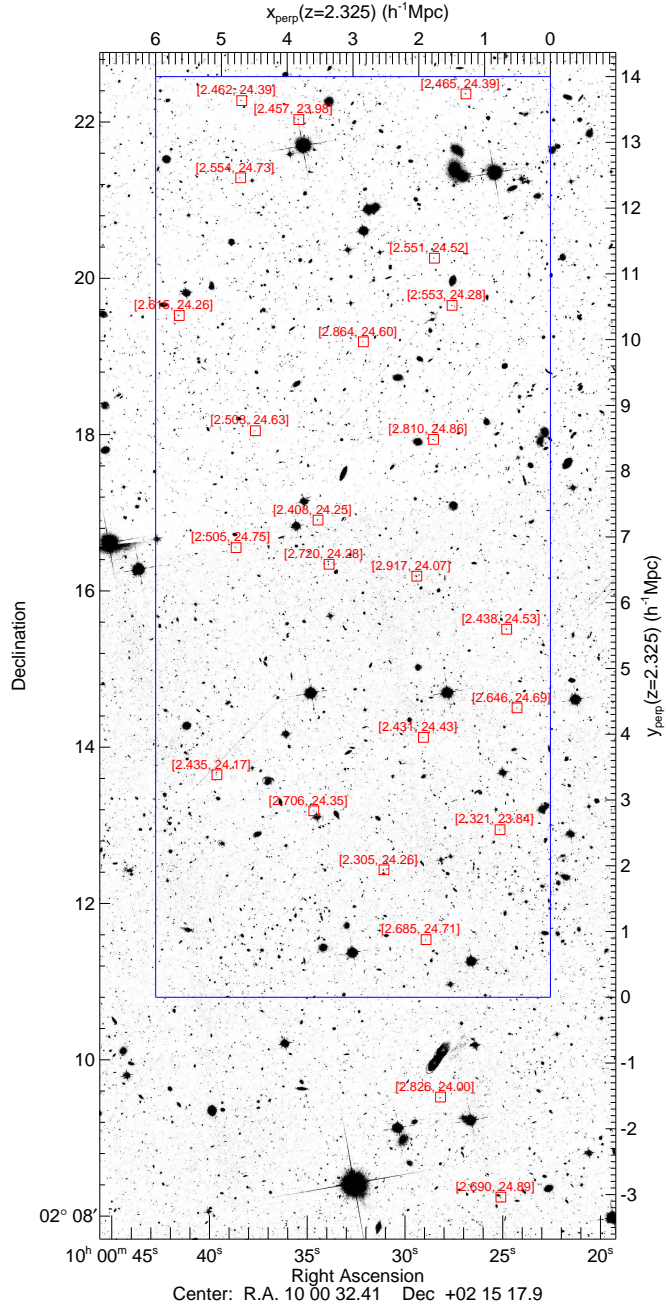}\caption{\label{fig:targets} 
%/Users/kheegan/lya/3d_recon/paper2/plotimg_targets.pro
HST ACS F814W mosaic \citep{koekemoer:2007} of our target region. 
The red boxes indicate our background spectroscopic sources with \lya\ forest coverage over $2.15\leq\zalp\leq2.40$;
source redshifts and $g$-magnitudes are labeled above each object. The transverse area of our map is 
bounded in blue; upper- and right-axes indicate the 
transverse comoving separation at $z=2.325$ relative to the map coordinate origins.}\end{center}\end{figure*}

Our observations target $g$-selected galaxies and AGN (using the \citealt{capak:2007} photometry) at
$2.3<\zbg<3$, such that their \lya\ forest
absorption covers $2.15\leq\zalp\leq2.40$. 
By working in the COSMOS field \citep{scoville:2007}, we are able to exploit rich multiwavelength
imaging and spectroscopy to efficiently target the necessary background sources.
Our primary candidates are spectroscopically-confirmed objects from the
 zCOSMOS-Deep \citep{lilly:2007} and VUDS \citep{le-fevre:2014} surveys ---
we reobserve these to obtain adequate S/N and spectral resolution for tomography. 
Where available, we also added grism redshifts kindly provided by
the 3D-HST team \citep[e.g.][]{brammer:2012}. 
Beyond spectroscopically-confirmed candidates, we add
 photometric redshifts from \citet{ilbert:2009} as well as
\citet{salvato:2011} for X-ray detected sources.
From these candidates, we selected targets based on redshift probability, source brightness, 
and uniformity on the sky --- 
the selection functions of the source catalogs are unimportant to us since the background source properties
do not bias the foreground absorption.

We observed with the LRIS Double-Spectrograph \citep{oke:1995,steidel:2004} on the 
Keck-I telescope at Mauna Kea, Hawai'i, 
during 2014 March 26-27 and 29-30, in MOS mode 
with the B600/4000 grism on the blue arm and R600/5000 grating on the red with 
the d500 dichroic. With $1\arcsec$ slits, this yields $R\equiv\lambda/\Delta\lambda\approx1000$ and $R\approx1200$ for the
blue and red arms, respectively. We suffered a $\sim 70\%$ weather loss, 
but obtained good-quality spectra for 2 
slitmasks covering $\sim 5\arcmin\times15\arcmin$, 
with total exposure times of 6600-7200s in $0.5-0.7\arcsec$ seeing and clear conditions. 
These two slitmasks overlap along their short edge, resulting in 
an elongated footprint (Figure~\ref{fig:targets}).

The data was reduced with the XIDL  
package\footnote{\url{http://www.ucolick.org/~xavier/LowRedux/lris_cook.html}}, and visually-inspected
to determine source redshifts. Out of 47 targeted objects, we successfully extracted 1D spectra and estimated redshifts for 33, 
of which 24 were determined to have the correct redshift and adequate
spectral S/N to contribute to our tomographic reconstruction.
The number of sources within our nominal $g \leq 24.5$ survey limit
 is $\sim 50\%$ that estimated by L14, a shortfall that was already evident during
the target-selection process. This is likely because L14 did not take into account 
dust-reddening \citep[$E(B-V)\sim0.2$,][]{reddy:2008} when estimating source counts ---
the SFG luminosity function is so steep that even small errors in the assumed magnitudes could easily
lead to this $\sim 50\%$ discrepancy. 
To fill the slitmasks, we therefore also targeted $g>24.5$ objects but these were less likely to be
successfully reduced or have adequate S/N. 
Nevertheless, even this reduced number of sources is sufficient to carry out
\lya\ forest tomography, as we shall see.

\begin{figure*}\begin{center}\includegraphics[width=0.7\textwidth,clip=true,trim=22 0 18 0]{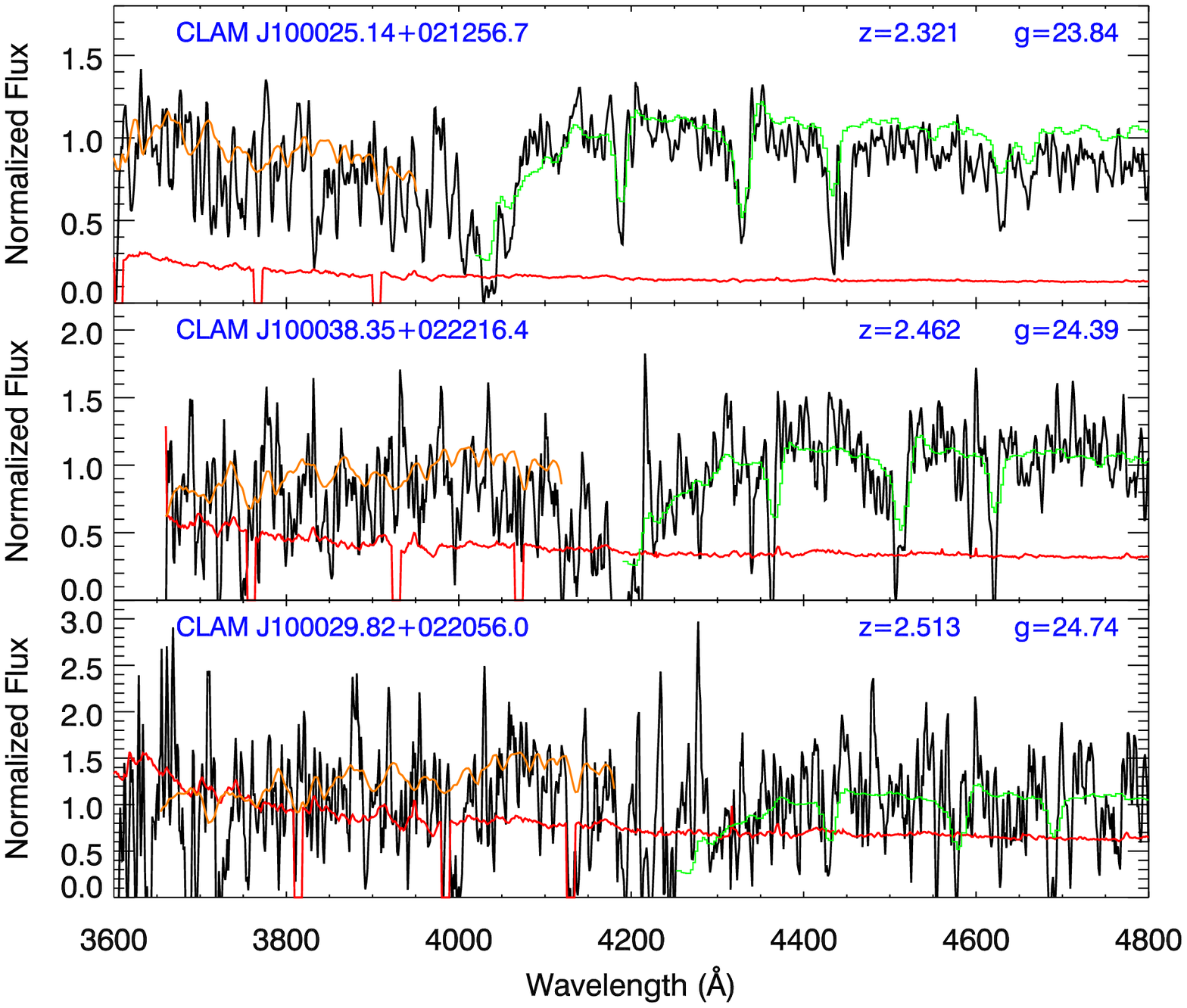}\caption{\label{fig:spectra}
Examples of SFG spectra obtained with Keck-LRIS 
and subsequently used for \lya\ forest tomographic reconstruction.
From top to bottom, these represent our highest-, median-, and lowest-S/N spectra, respectively.  
The red curve represents the estimated pixel noise, with masked pixels (mostly intrinsic absorption-lines) set to zero. 
The green curve is the \citet{shapley:2003} 
composite LBG spectrum overplotted at the source redshifts, while the orange curve is the estimated continuum 
(see text).}\end{center}\end{figure*}

The position of the 24 SFGs on the sky are shown in Figure~\ref{fig:targets}.
%KG: Omitted to save words
%(one object, CLAM J100024.80+021530.6, was in fact observed in both masks). 
%None of the spectra exhibit broad emission-lines indicating AGN activity; 
Our brightest objects 
 are $g \approx 24.0$ SFGs with $\snr \approx 3-4$ per $1.2\,\ang$ pixel,
while on the faint-end we use spectra down to $\snr \approx 1.3$ from $g\approx 24.8$ sources. 
Examples of the spectra are shown in Figure~\ref{fig:spectra}. 
We also attempted to visually identify damped \lya\ absorbers that might
 affect \lya\ forest analysis but found none.% --- while it's possible that we have missed DLAs in the noisier spectra,
 %such noisy spectra are downweighted in our analysis and any DLA contamination is unlikely to significantly
% bias the resulting map.
 
 % spectra/plot_lbg_mosaic.pro
 
 To extract the \lya\ forest transmission from the spectra, 
 we need to estimate the intrinsic `continuum' of the SFGs. 
 Studies of $z\sim3$ SFG composite spectra  \citep{shapley:2003,berry:2012} suggest that this is relatively
 flat in the \lya\ forest region, with only a few strong intrinsic absorbers visible --- this is corroborated by
 high-resolution line
 analysis of the lensed galaxy MS1512-cB58 \citep{savaglio:2002}.
 From these studies, we determined that the strongest intrinsic absorption
 within the $1040-1190\,\ang$ \lya\ forest region are at
 \waveion{N}{2}{1084.0}, \waveion{N}{1}{1134.4}, and \waveion{C}{3}{1175.7} --- we mask $\pm5\,\ang$ 
 around these transitions.  
 We then adopt as our continuum template the restframe composite spectrum of 59 SFGs from \citet{berry:2012}, in which the \lya\ forest
 variance in the restframe 
 $\sim 1040-1190\,\ang$ region have been smoothed out through averaging, albeit with an overall absorption 
 decrement.
 %\footnote{The \citealt{shapley:2003} composite was constructed from more objects, 
 %but the spectral resolution ($R\sim 400$) is insufficient for our $R\sim 1000$ data.}. 
 
 Using this template, we estimate the continuum, $C$, for each individual spectrum by `mean-flux regulation' 
\citep{lee:2012a,lee:2013}, i.e. adjusting the amplitude and slope of the $1040-1190\,\ang$ continuum template 
until the mean \lya\ 
%% JFH Please cite your BOSS PDF paper here. We need to cite it and this is as good a place as any. 
forest transmission, 
$\fmean(z)$, from each spectrum agrees with the measurements of \citet{becker:2013}. 
This method ensures that there is no overall bias in the resulting continua.  
We estimate the continuum error to be $\lesssim10\%$, by considering the variation of
Starburst99 \citep{leitherer:1999, leitherer:2010} models with respect to various physical parameters.
This is adequate for our $\snr\leq4$ spectra, but in future papers we will study SFG continuum-fitting in more detail.

We divide the restframe $1040-1190\ang$ flux, $f$, from each spectrum by the continuum to obtain the \lya\ forest
transmission $F=f/C$, and further the forest fluctuations:
\beq\delta_F=F/\langle F\rangle(z)-1.\eeq 
We also compute the error, $\sigma_N=\sigma/C/\langle F\rangle(z)$, where $\sigma$ is 
the pixel noise reported by the reduction pipeline. The vectors of $\delta_F$ and $\sigma_N$, along with the corresponding
3D pixel positions, constitute the inputs for the tomographic reconstruction.

\section{Tomographic Reconstruction}

\begin{figure*}[t!]\begin{center}{\includegraphics[scale=0.33,trim=0 0 10 20]{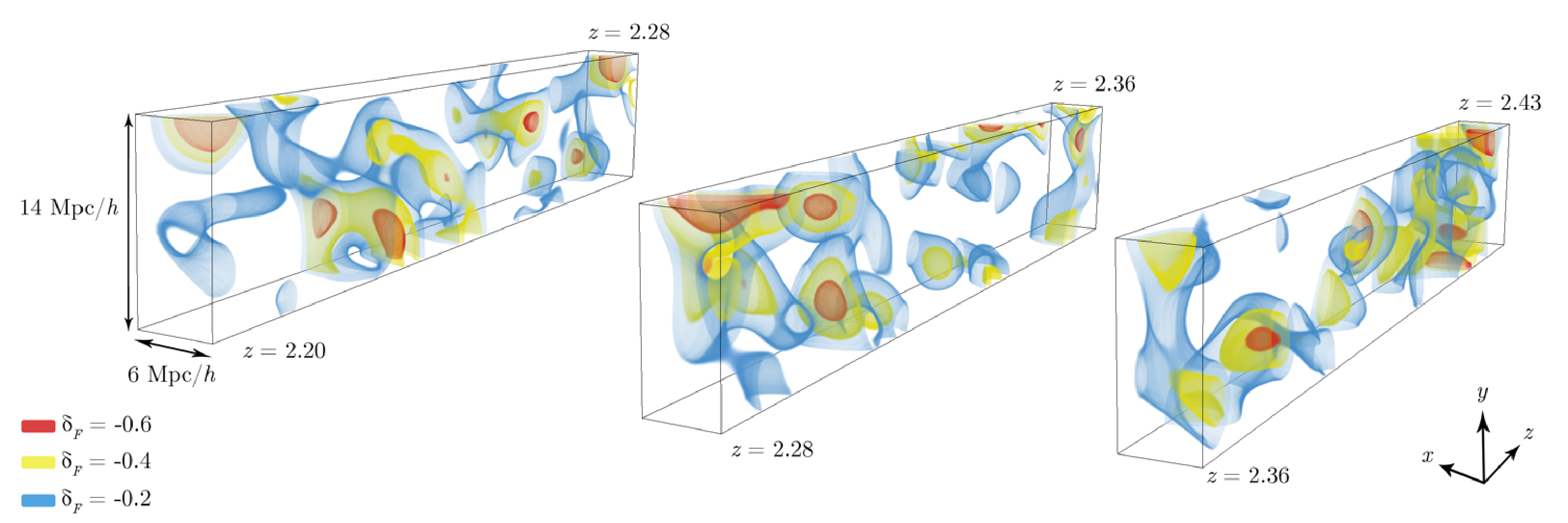}}\\{\includegraphics[scale=0.39,clip=true,trim=8 15 10 0]{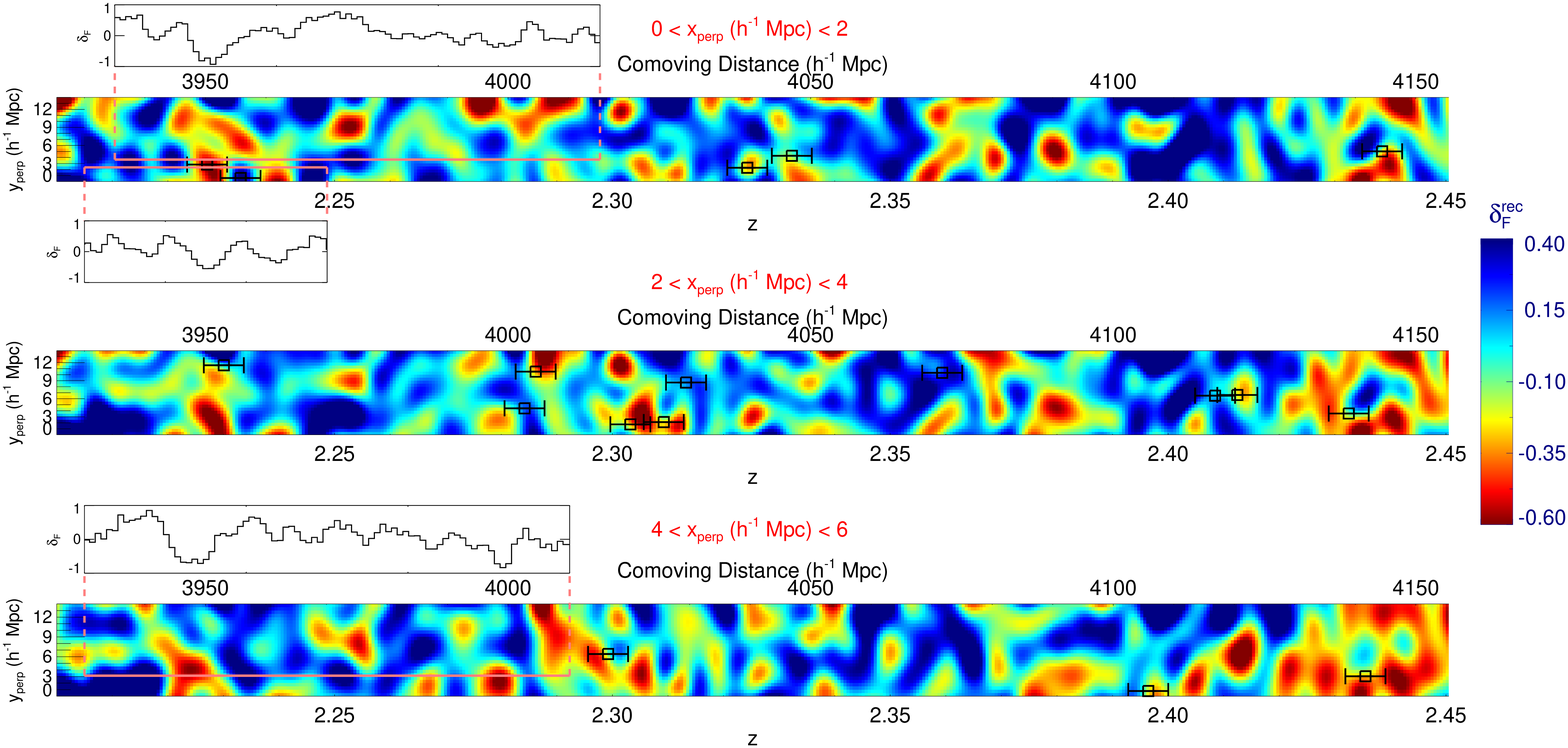}}
\caption{\label{fig:mapslice} Tomographic reconstruction of 3D \lya\ forest absorption from our data, 
shown in 3 redshift segments in 3D (top) and projected over 3 slices along the R.A. direction 
(bottom panels).
The color scale represents reconstructed \lya\ forest transmission such that negative values (red) correspond to overdensities. Square symbols denote positions of coeval galaxies within the map; 
error bars indicate the $\sigma_v\approx300\,\kms$ uncertainty on their redshifts. Pink solid lines
indicate where 3 of the skewers probe the volume, with inset panels indicating the corresponding 1D absorption spectra (top-hat smoothed by 3 pixels) that contributed to the tomographic reconstruction.}\end{center}\end{figure*}

To create the \lya\ forest tomographic reconstruction, 
we use Wiener filtering \citep[e.g.,][]{wiener:1942,press:1992,zaroubi:1995}, 
where the reconstructed field, $\delrecon$, is:
\beq \label{eq:wiener}\delrecon=\cmd\cdot (\cdd+\mathbf{N})^{-1}\cdot\delta_F,\eeq
where $\cdd+\mathbf{N}$ and $\cmd$ are the data-data and map-data covariances, respectively. 
The noise covariance matrix $\mathbf{N}$ is assumed to have only diagonal elements set by the noise variances, 
$N_{ii}=\sigma_{N,i}^2$. 
This term allows us to weight each input pixel by its S/N, so lower-S/N spectra
are down-weighted and avoids noise spikes from biasing the map.

Following L14 and \citet{caucci:2008}, we assume that between any two points $\mathbf{r_1}$ and $\mathbf{r_2}$,
whether in the maps or skewers,  $\cdd=\cmd=\mathbf{C(r_1,r_2)}$ and 
\beq\mathbf{C(r_1,r_2)}=\sigma_F^2\exp\left[-\frac{(\Delta r_\parallel)^2}{2L^2_\parallel}\right] \exp\left[-\frac{(\Delta r_\perp)^2}{2L^2_\perp}\right],\eeq
where $\Delta r_\parallel$ and $\Delta r_\perp$ are the distance between 
$\mathbf{r_1}$ and $\mathbf{r_2}$ along, and transverse to. the line-of-sight, respectively. 
$L_\parallel$ and $L_\perp$ are free parameters that 
set the effective smoothing of the reconstruction parallel and perpendicular to the line-of-sight, respectively,
while $\sigma_F=0.8$ sets the overall correlation strength.
These parameters need to be matched to the data quality:  
we set $L_\parallel=2.7\,\hMpc$, roughly the comoving scale along the LOS corresponding 
to our spectral resolution element. For $L_\perp$, \citet{caucci:2008} suggested 
setting it to the typical transverse sightline separation $\dperp$, but
 we choose $L_\perp=3.5\,\hMpc$ even though our sightline separation is
$\dperp\approx 2.3\,\hMpc$. 
This is a conservative choice taking into account the low-S/N of our individual spectra.
The choice of these reconstruction parameters is somewhat arbitrary since small changes do not qualitiatively
change the resulting map features, but in future work we will discuss optimal choices for these parameters.

% sims/plot_mocks_slice.pro
\begin{figure*}[t!]\begin{center}\includegraphics[height=0.5\textheight]{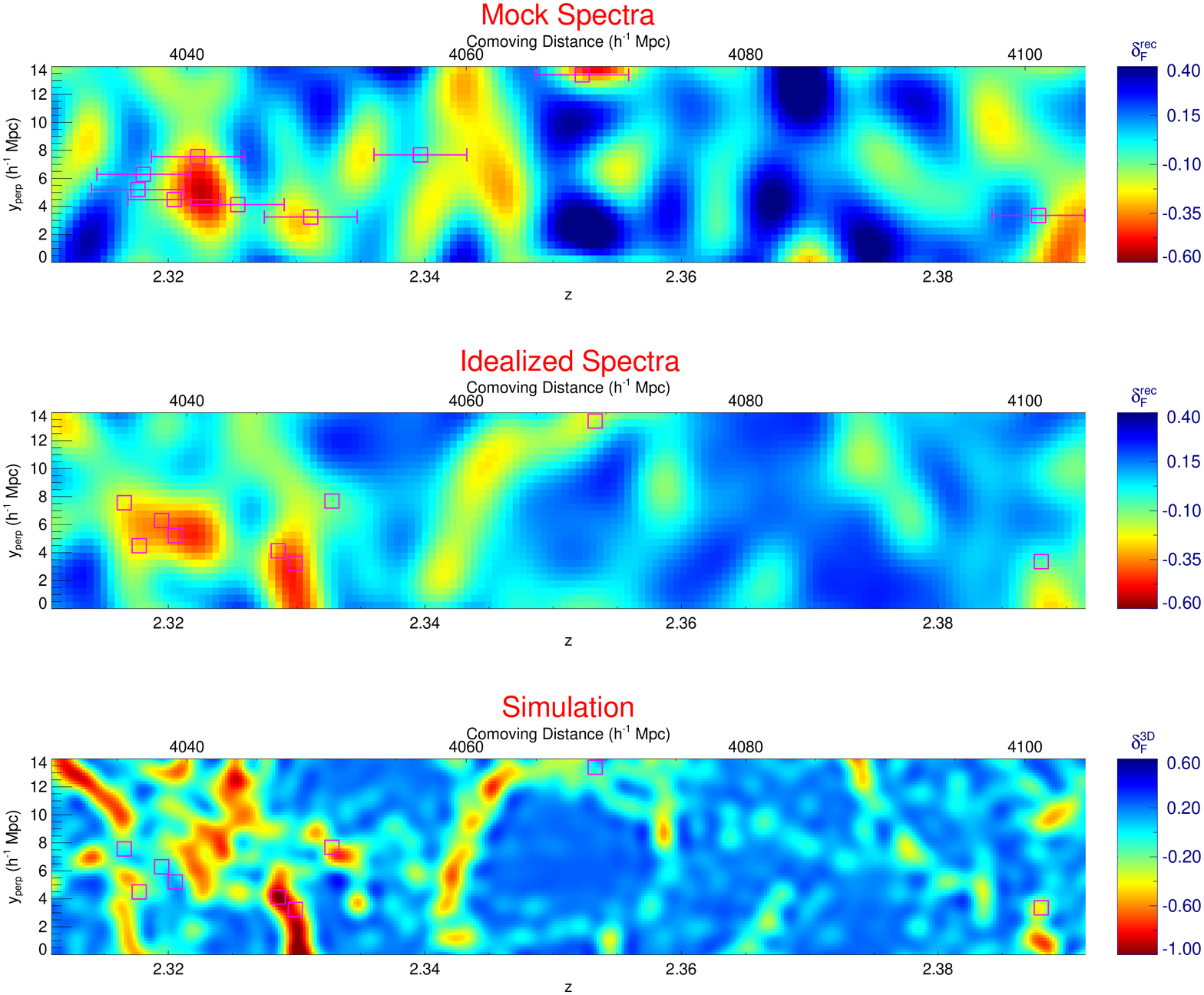}
\caption{\label{fig:mocks} 
(Top) A slice from a tomographic reconstruction (projected over $\Delta\xperp=2\,\hMpc$) 
using a mock data set with similar spatial sampling and S/N to our data.
(Middle) A reconstruction (with the same $[L_\parallel,L_\perp,\sigma_F]$) from the 
full grid of noiseless spectra with
$0.8\,\hMpc$ transverse separations.
For reference, the bottom panel shows the `true' absorption field in the simulation. 
Magenta squares indicate locations of coeval $R\leq 25.5$ galaxies --- in the top panel 
we also introduced random redshift errors.}\end{center}\end{figure*}

Our map originates at $[\alpha_0,\delta_0]=[10^{h}00\arcmin22\fs56,+02\degr10\arcmin48.0\arcsec]$, spanning $[6\hMpc,14\hMpc]$ 
in the $[x_\mathrm{perp}, y_\mathrm{perp}]$ directions on the sky (c.f. top- and right-axes in Figure~\ref{fig:targets}); 
along the line-of-sight, the origin is $\zalp=2.20$ and extends $\Delta\chi=230\,\hMpc$ up to 
$\zalp\approx2.45$, giving an overall comoving volume of 
$6\,\hMpc\times14\,\hMpc\times230\,\hMpc=19320 h^{-3}\,\mathrm{Mpc}^3\approx(27\,\hMpc)^3$. 
Note that our map does not cover the region $\delta\lesssim 2\degr11\arcmin$, 
where we experienced a high failure-rate in spectral-extraction and redshift-identification due to 
deteriorating observing conditions.
However, the two spectra in the excluded region are still included in the map input; given our 
transverse correlation length of $L_\perp=3.5\,\hMpc$, these spectra 
($\approx 1.5\,\hMpc$ and $\approx 3\,\hMpc$ from the lower map boundary)
still contribute to the low-$y_\mathrm{perp}$ portions of the map.

We evaluated Equation~\ref{eq:wiener} to solve for the output tomographic map, $\delrecon$, 
using a preconditioned conjugate-gradient
algorithm to carry out the matrix inversion and matrix-vector multiplication (C.~Stark et al., in preparation),
sampling on a 3D comoving grid with $(0.5\,\hMpc)^3$ cells.
 For simplicity, we assumed a fixed differential comoving distance $d\chi/dz$ (evaluated at $z=2.325$, the
 mean map redshift) when setting up the output grid. 
This avoids a flared map geometry, since the transverse comoving area increases with 
redshift, but within our limited redshift range this effect is small.
  
The resulting map of the 3D \lya\ forest absorption, $\delrecon$, 
is shown in Figure~\ref{fig:mapslice} as 3D visualizations and slices projected over the $\xperp$ (R.A.) direction.
A lot of structure is obvious even
within this small volume, with overdensities (negative-$\delrecon$ regions) spanning comoving distances of 
$\Delta y_\mathrm{perp}\gtrsim10\,\hMpc$ both along the line-of-sight (e.g. from $z\approx 2.21$ to $z\approx 2.23$ at 
$\yperp\sim8\,\hMpc$) and across the transverse direction (at $z\approx 2.43$).  
The strong overdensities are typically sampled by multiple sightlines
at different background redshifts. This is illustrated by inset panels in the map slices in 
Figure~\ref{fig:mapslice}, where we show 3 examples of the 1D absorption field, $\delta_F$, that went into
the reconstruction --- the overdensity at a comoving distance of $\approx 3950\,\hMpc$ 
and $\yperp \leq 5\,\hMpc$ can be seen as clear dips
in all three of the spectra, which is unlikely to be caused by pixel noise. Note that in moderate-resolution \lya\ forest data, 
 significant `absorbers' are typically due to blends of clustered \lya\ forest absorption and not 
individual absorbers \citep{lee:2014b,pieri:2013}. 
One also clearly sees significant voids (dark blue regions) on scales of $\sim5-10\,\hMpc$.

As validation, we performed reconstructions on mock data sets derived from simulations (e.g., L14). 
These mocks have identical sightline configurations, 
resolution, and S/N as the data, including 
random continuum errors with 7\% RMS.
The resulting reconstructions are illustrated in Figure~\ref{fig:mocks}, compared
with the `true' absorption field from the simulation. 
The good correspondence between large-scale features in the `true' and reconstructed fields gives us 
confidence that the real map (Figure~\ref{fig:mapslice}) is indeed probing LSS.
However, the PDF of the simulated reconstructions differed
 from the real map (c.f. black histograms in Figures~\ref{fig:maphist}a and b).
To investigate, we ran 24 mock reconstructions on independent simulation volumes, 
which showed considerable scatter in the resulting PDFs (shaded grey area in Figure~\ref{fig:maphist}b).
This suggests that part of the discrepancy is due to cosmic-variance from our small volume.
Moreover, while DM-only simulations correctly reproduce \lya\ forest clustering, 
they do not yield the right PDF \citep{white:2010},
which could also contribute to the disagreement.

L14 argue (e.g.\ their Figure~6) that the reconstructed $\delrecon$ scales  
approximately linearly with the dark-matter overdensity, 
$\Delta_\mathrm{dm}\equiv\rho_\mathrm{dm}/\langle\rho_\mathrm{dm}\rangle$, smoothed on similar scales
($\sigthreed\approx3.5\,\hMpc$ in our case), albeit with some scatter due to reconstruction noise.
 The most negative $\delrecon$ correspond to overdensites 
 of $\Delta_\mathrm{dm}\approx2$ while the most positive $\delrecon$ indicate underdensities of $\Delta_\mathrm{dm}\approx0.2$.

\section{Comparison with Coeval Galaxies}
%compare_galaxies.pro
%compare_galaxies_sim_mock16.pro
%compare_galaxies_perc.pro
\begin{figure*}[t!]\begin{center}\includegraphics[width=0.48\textwidth]{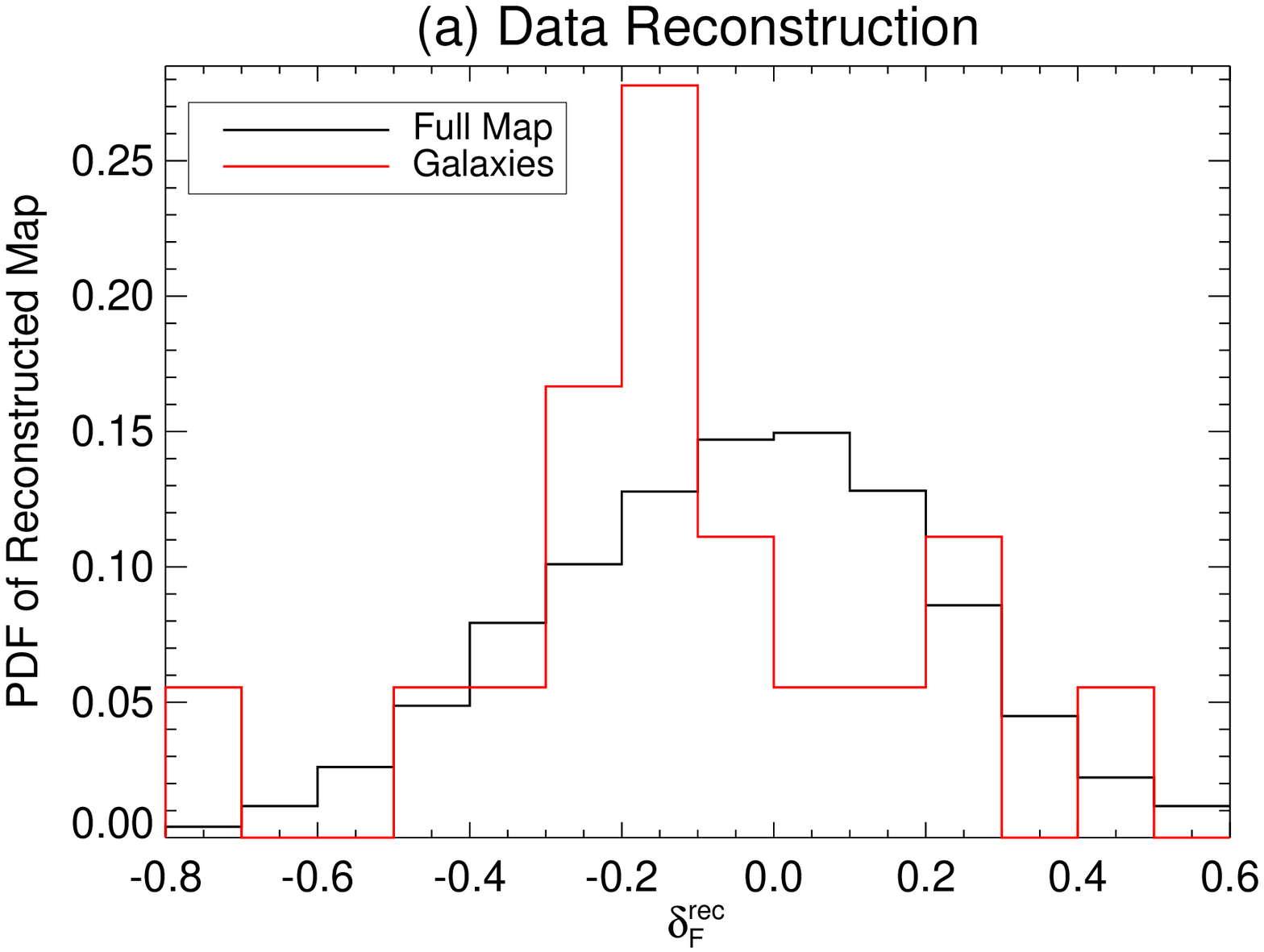}\hspace{2em}\includegraphics[width=0.48\textwidth]{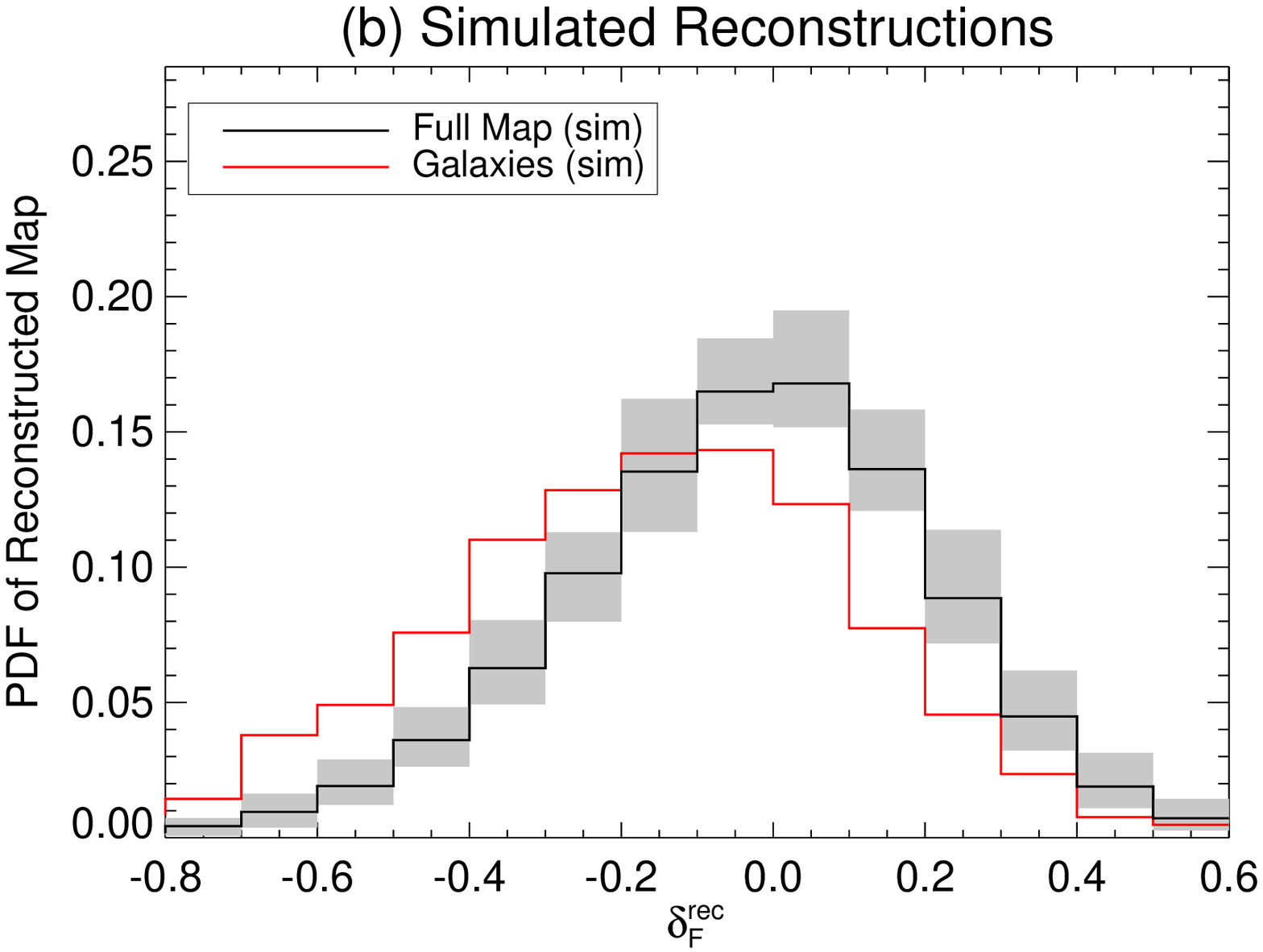}\\ \vspace{0.5em}\includegraphics[width=0.48\textwidth]{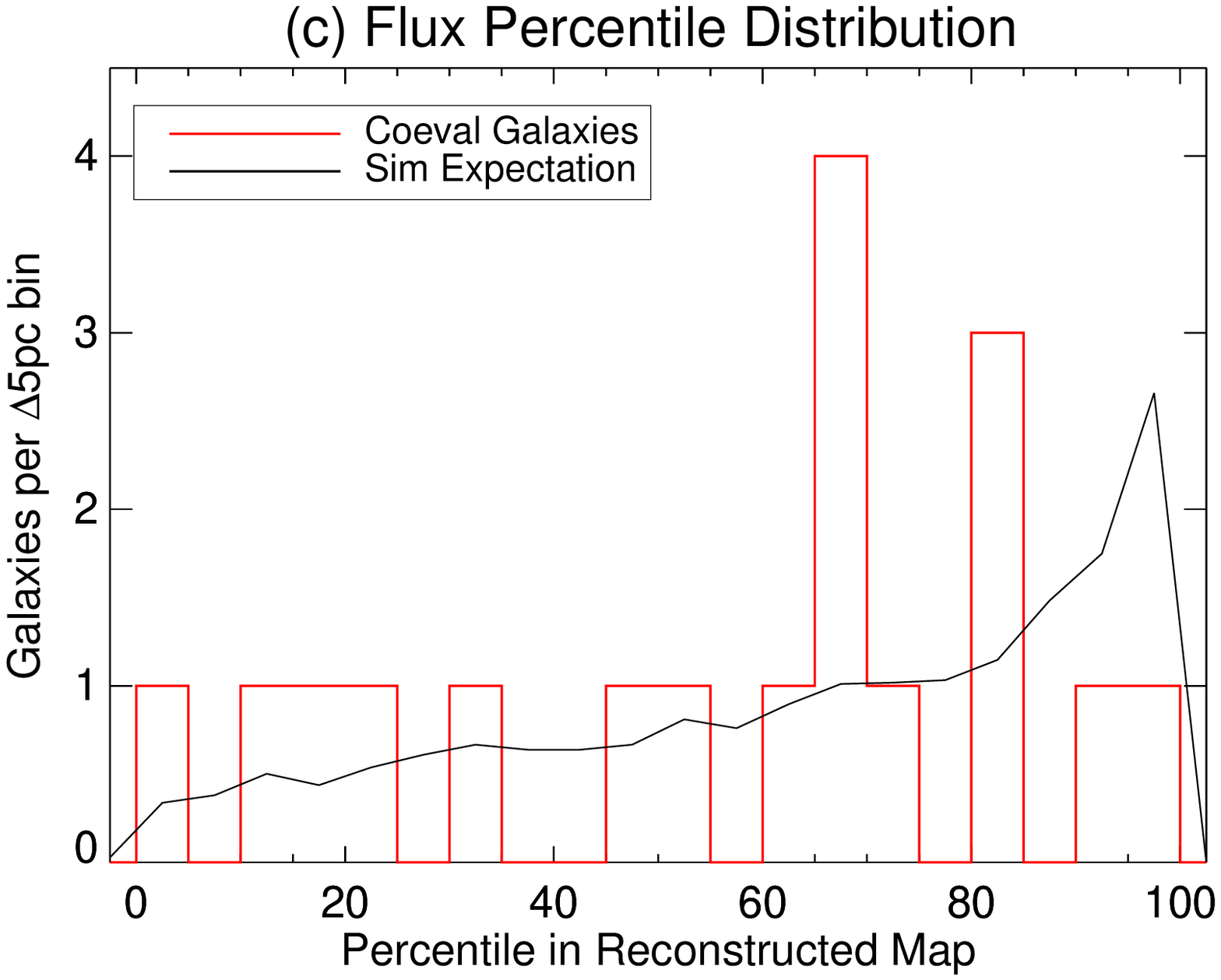}\end{center}\caption{\label{fig:maphist} 
(a) PDF of our 
tomographic map (black) compared with that sampled by 18 coeval galaxies within our map volume
(red, both PDFs normalized to unit area). 
(b) Similar to (a), but evaluated over 24 mock reconstructions simulating the real map. 
The red curve shows the $\delrecon$ evaluated at 2506 simulated $R\leq25.5$ galaxies within the mock reconstructions --- the simulated galaxies
clearly also preferentially live in low-$\delrecon$ regions. 
Shaded regions indicate
the range of map PDFs from the 24 mock reconstructions, indicating the significant sample variance from the small
volume.
(c) Distribution of coeval galaxies as a function of the
map flux percentile, such that $\delrecon$ decreases with the percentile, i.e. larger percentiles probe overdensities. 
The black curve indicates
the predicted distribution
from the simulated galaxies within our mock reconstructions.}\end{figure*}

% REDRUM REDRUM REDRUM

Since galaxies are well-known tracers of LSS, we can exploit the 
spectroscopically-confirmed high-redshift galaxies within the COSMOS field 
\citep{lilly:2007,le-fevre:2014} to make a comparison with our \lya\ forest tomographic map.
We searched an internal COSMOS compilation of all available spectroscopic redshifts, 
and found 18 galaxies coeval within 
the map volume (4 were uniquely confirmed by our observations).
This small number is clearly inadequate for mapping $z\gtrsim2$ LSS on $\sim\mathrm{Mpc}$ scales,
illustrating the challenge of using galaxy redshift surveys for this purpose, despite many hundred hours of large-telescope time.
In order to make galaxy maps with comparable resolution to our tomographic reconstructions, 
the galaxy number density needs to be increased dramatically, requiring 30m-class telescopes to obtain redshifts
from faint ($R\gtrsim 26$) galaxies.
  
For these coeval galaxies, we determined their 3D positions within our map 
(overplotted on Figure~\ref{fig:mapslice})
 and evaluated the corresponding $\delrecon$.  
The $\delrecon$ sampled by these galaxies are shown in  
Figure~\ref{fig:maphist}a, compared with the $\delrecon$ distribution from the full map;
in Figure~\ref{fig:maphist}c, we show the galaxy distribution as a function of the percentile of map ranked by
flux (where larger flux percentiles represent overdensities). 
The galaxies preferentially occupy low-$\delrecon$ regions (i.e. overdensities) of the map.

However, at first glance it seems troublesome that several galaxies are located in high-$\delrecon$ (underdense) regions. 
This could partly be due to errors in 
the galaxy spectroscopic redshifts: these are $\sigma_v\approx 300\,\kms$ \citep{diener:2013}, 
i.e. $\sigma_\chi\approx3.3\,\hMpc$ along the LOS at $z\approx2.3$.
This seems plausible for the galaxy at $[\xperp, \yperp, z]=[0.5\,\hMpc,\,0.6\,\hMpc,\, 2.233]$ (top panel, Figure~\ref{fig:mapslice}),
which apparently
occupies a void but is in fact within $\pm1\sigma$ of two overdensities on either side.
Indeed, in Figure~\ref{fig:mapslice} most of the galaxies are within $\sim 1\sigma$ of significant overdensities.
Another possible reason for this discrepancy could be different redshift-space distortions experienced by
the galaxies and the forest: the latter has been constrained by \citet{slosar:2011} but yet to be measured for 
$z\gtrsim2$ galaxies.

Tomographic reconstruction errors (c.f., Figure~\ref{fig:mocks}) could also decrease the correlation between 
the galaxies and 3D \lya\ absorption, particularly in regions poorly-sampled by sightlines. 
We investigate this using our simulations, from which we extracted $R\leq25.5$ galaxies
through halo abundance-matching (see L14 for details), 
introduced the expected LOS redshift errors and then evaluated their positions
within the mock tomographic reconstructions; this is illustrated by the mock galaxies in Figure~\ref{fig:mocks}. 
The distribution is shown in the red histogram in Figure~\ref{fig:maphist}b, 
which shows a clear preference towards negative-$\delrecon$ (overdensities). This is also evident in Figure~\ref{fig:maphist}c, 
which shows the distribution as a function of flux percentiles (normalized to $N=18$ as in the real data). 
A two-sample Kolmogorov-Smirnov 
test between the percentile distribution of the real galaxies versus that from the simulations indicate
22\% probability of being drawn from the same distribution, which is reasonable considering the small data set.
The long tail of galaxies in the underdensities  
is primarily due to a combination of galaxy redshift errors and reconstruction noise.
The former could be mitigated in the near-future by accurate systemic redshifts from near-IR spectroscopy,
while to account for reconstruction noise 
we are developing methods to estimate the 
map covariance and hence characterize the uncertainties at any point
within the maps.

\section{Conclusion}
We present the spectroscopic observations targeting, for the first time, 
 high-redshift galaxies as background sources for \lya\ forest analysis. 
This enabled us to create a tomographic map of the 3D absorption field with a spatial resolution of 
$\sigthreed\approx3.5\,\hMpc$ covering a comoving volume of 
$\approx (27\,\hMpc)^3$ at $\langle z\rangle\approx2.3$. 
Simulated tomographic reconstructions show that our sightline-sampling, resolution, and 
S/N should yield a good recovery of the underlying absorption field.
Supporting this conclusion, a sample of 18 coeval galaxies with
known spectroscopic redshifts are found to
preferentially occupy high-absorption regions (i.e. overdensities) in our map.

These results demonstrate the feasibility and promise of
the full CLAMATO survey: $\sim1000$ SFGs at $\zbg\sim2-3$ 
covering $\sim1\,\sqdeg$ in the COSMOS field, which will enable a $\langle z\rangle\sim2.3$ 
\lya\ forest tomographic map with $\sigthreed\sim3-4\,\hMpc$ spatial resolution over a 
$(65\,\hMpc)^2 \times250\,\hMpc\sim(100\,\hMpc)^3$ comoving volume.
This will allow us to directly characterize the topology and morphology of $z>2$ LSS for the first time
--- already we see tantalizing hints of structures extending across 
$\gtrsim10\,\hMpc$ in the high-redshift cosmic web.
A large-volume LSS map will also enable a search for
progenitors of massive $z\sim0$ galaxy clusters --- these protoclusters should manifest themselves at $z\gtrsim 2$ as
overdensities of a few over $\sim10\,\hMpc$ \citep{chiang:2013} scales. In a forthcoming paper, 
we will discuss methods to find protoclusters using \lya\ forest tomography.
 
The proposed survey will create rich synergy with other COSMOS datasets. 
We would be able to study various high-redshift galaxy 
properties, e.g., morphology, color, 
star-formation rate, as a function of their environment within the cosmic web. 
Such studies will require the full $\sim(100\,\hMpc)^3$ CLAMATO volume in order to sample enough 
objects to beat down the galaxies'
redshift uncertainties and reconstruction errors, 
but promises unique insights into galaxy formation and evolution during the $z\sim2-3$ epoch.
Finally, CLAMATO will probe small-scale clustering of LSS, and will be 
highly-complementary with wide-field surveys such as HETDEX \citep{hill:2004}
and DESI \citep{levi:2013} to probe cosmological clustering over a broad range of spatial-scales at $z\gtrsim2$.
  
\acknowledgements{
%We thank Eric Gawiser, Rupert Croft, Nao Suzuki, Arjen van der Wel, and Mike Maseda for fruitful discussions
%and comments. 
%KGL thanks David Spergel for humoring his crackpot ideas, and
%Michael Strauss for asking tough questions at his Ph.D. defence.
KGL and ACE are grateful to the National~Geographic Society for travel support through the Waitt Grants program.
This research used resources of the NERSC Center, which is supported by the Office of Science of the U.S. 
D.O.E. under Contract \#DE-AC02-05CH11231.
%The Keck observations described in this paper were conducted through the programs U047LA (P.I.: Schlegel) and S300LA 
%(P.I.: Lee). 
We would like to thank those of Hawai'ian ancestry, on whose sacred mountain 
we were privileged to be guests.
} \\

\bibliographystyle{apj}
%\bibliography{lyaf_kg,apj-jour,lss_galaxies}

\end{document}